\documentclass[twocolumn,prl,english]{revtex4-1}
\usepackage[T1]{fontenc}
\usepackage[latin9]{inputenc}
\usepackage{verbatim}
\usepackage{amsmath}
\usepackage{amssymb}
\usepackage{graphicx}
\usepackage{babel}
\usepackage{bm}
\usepackage[normalem]{ulem}
\usepackage{footnote}
\newcommand{\ubar}[1]{\mkern 1.5mu{\uline{{\mkern-1.5mu#1\mkern-1.5mu}}}\mkern 1.5mu}
\newcommand{\ubaro}[1]{\mkern 1.5mu\mbox{\uline{{$\mkern-1.5mu#1\mkern-1.5mu$}}}\mkern 1.5mu}
\newcommand{\uubar}[1]{\ubar{\ubaro{#1}}}

\newcommand\uuuline{\bgroup\markoverwith%
	{%
		\textcolor{black}{\rule[-0.5ex]{2pt}{0.4pt}}%
		\llap{\textcolor{black}{\rule[-0.8ex]{2pt}{0.4pt}}}%
		\llap{\textcolor{black}{\rule[-1.15ex]{2pt}{0.4pt}}}%
	}%
	\ULon}

\makeatother

\makeatletter

\usepackage{xcolor}
\newcommand*\diff{\mathop{}\!\mathrm{d}}

\newcommand{\SC}[1]{\textcolor{black} {#1}}
\newcommand{\be}{\begin{equation}}
\newcommand{\ee}{\end{equation}}

\begin{document}
\renewcommand{\ULdepth}{1pt}
\title{
Field theory for mechanical criticality in disordered fiber networks}

\author{Sihan Chen$^{1,2,3,4}$, Tomer Markovich$^{2,5,6}$, Fred C. MacKintosh$^{1,2,7,8,9}$}
\affiliation{$^1$Department of Physics and Astronomy, Rice University, Houston, TX 77005, US\\
	$^2$Center for Theoretical Biological Physics, Rice University, Houston, TX 77005, US\\
	$^3$Kadanoff Center for Theoretical Physics, University of Chicago, Chicago, IL 60637, US\\
	$^4$The James Franck Institute, University of Chicago, Chicago, IL 60637, US\\
	$^5$School of Mechanical Engineering, Tel Aviv University, Tel Aviv 69978, Israel\\
	$^6$Center for Physics and Chemistry of Living Systems, Tel Aviv University, Tel Aviv 69978, Israel\\
	$^7$Department of Chemical and Biomolecular Engineering, Rice University, Houston, TX 77005\\
	$^8$Department of Chemistry, Rice University, Houston, TX 77005\\
	$^9$The Isaac Newton Institute for Mathematical Sciences, Cambridge University, Cambridge, UK}

\begin{abstract}
Strain-controlled criticality governs the elasticity of jamming and fiber networks. While the upper critical dimension of jamming is believed to be $d_u=2$, non-mean-field exponents are observed in numerical studies of 2D and 3D fiber networks. The origins of this remains unclear. In this study we propose a minimal mean-field model for strain-controlled criticality of fiber networks. We then extend this to a phenomenological field theory, in which non-mean-field behavior emerges as a result of the disorder in the network structure. We predict that the upper critical dimension for such systems is $d_u=4$ using a Gaussian approximation.  Moreover, we identify an order parameter for the phase transition, which has been lacking for fiber networks to date.  
\end{abstract}

\maketitle
Strain-controlled rigidity transitions and criticality have been identified in systems ranging from the jamming of particle suspensions to fiber networks and extracellular matrices~\cite{o2002random,wyart2008elasticity,liu2010jamming,goodrich2016scaling,Sharma2016,jansen2018role}. These systems demonstrate a transition from a floppy to a rigid phase with increasing applied strain.  Similar physics also appears to govern shear thickening as a function of strain rate~\cite{wyart2014discontinuous,
ramaswamy2023universal,
chacko2018dynamic,
hertaeg2022discontinuous}.
Both fiber networks and suspensions exhibit striking features of critical phenomena near the onset of rigidity, including power-law distributed forces~\cite{hagh2019broader}, diverging non-affinity~\cite{arevalo2017nonaffinity,van2009jamming,shivers2019scaling,ikeda2021nonaffine,chen2023nonaffine} and critical slowing-down in dynamics~\cite{atkinson2016critical,shivers2022nonaffinity}. Despite the similarities between jamming and fiber networks, however, a notable distinction lies in the nature of their critical exponents. While frictionless jamming is mean-field in dimension $d\geq 2$~\cite{goodrich2016scaling,sartor2021mean}, non-mean-field exponents have been reported for both 2D and 3D fiber networks~\cite{Sharma2016,sharma2016strain,vermeulen2017geometry,shivers2019scaling,shivers2020nonlinear,arzash2020finite,arzash2021shear,arzash2022mechanics,lee2022stiffening}. These non-mean-field exponents in fiber networks hint at an underlying difference in the nature of the two transitions, and a theoretical understanding of this criticality is lacking. 

Here, we present a phenomenological field theory for strain-controlled criticality of fiber networks. We first propose a minimal model that reproduces the mean-field behavior of the phase transition. We then extend this to a field theory that incorporates network disorder, with which we are able to calculate an anomalous critical exponent and identify the upper critical dimension $d_u=4$. We show how reduced levels of disorder, and particularly hyper-uniformity~\cite{torquato2003local,hexner2015,wilken2020hyperuniform} with vanishing long-wavelength fluctuations leads to mean-field behavior.  
Based on this theory, we also propose and computationally verify an order parameter for this transition, which has been lacking to date.

\begin{figure}[t]
	\centering
	\includegraphics[scale=0.35]{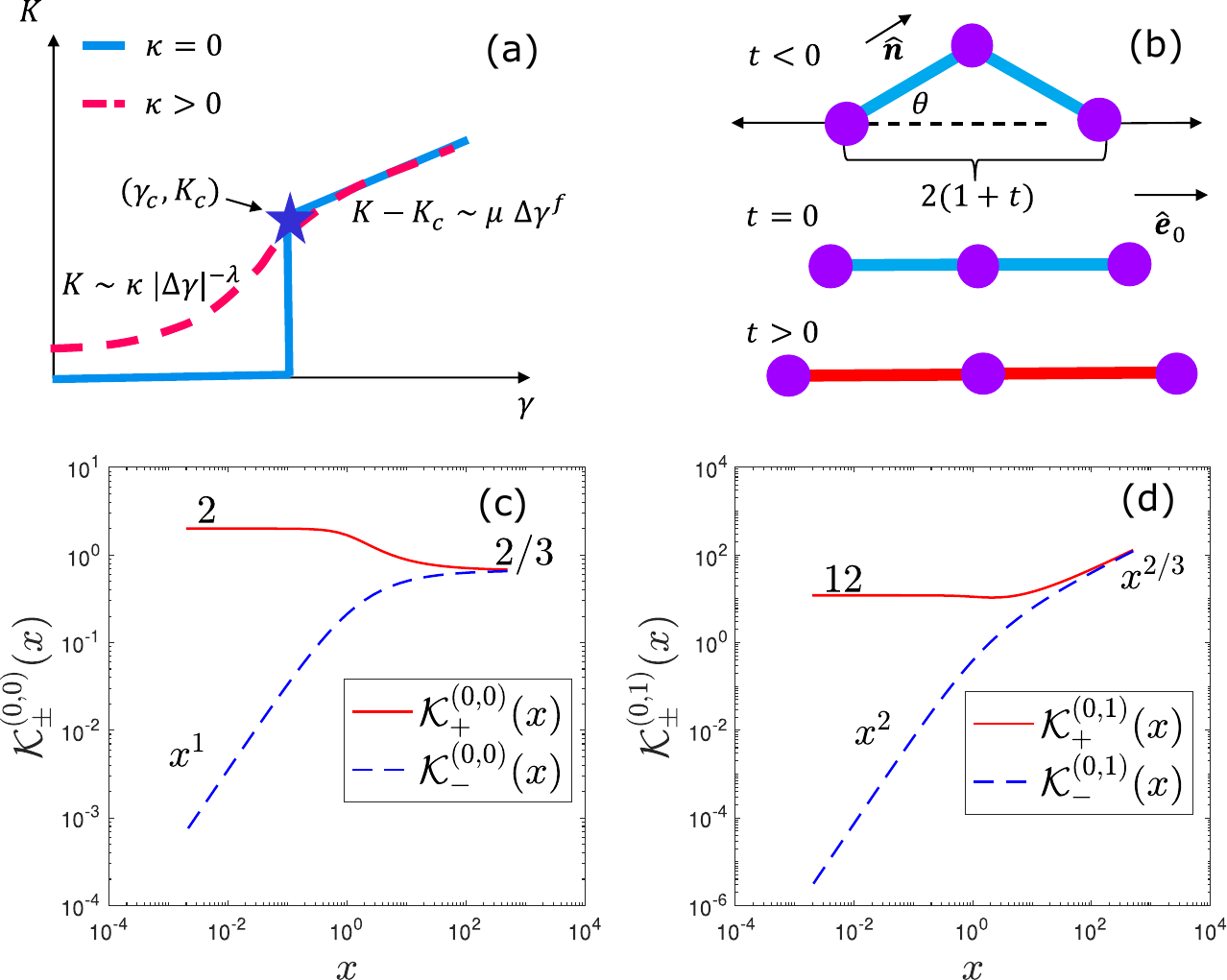}
	\caption{(a) Schematic diagram of the scaling behavior of the stiffness $K$ of fiber networks. (b) Minimal model for strain-controlled phase transition, formed by two connected segments under an extensile strain in the direction $\hat{\bm e}_0$. Red indicates tense segments for $t>0$.  (c) and (d) Scaling functions of the elasticity for the minimal model. 
	}
	\label{fig_1}
\end{figure}
\textit{Strain-controlled criticality} \textendash
We start by briefly summarizing previous observations on strain-controlled criticality of fiber networks. Fiber networks are modeled as athermal networks of interconnected segments~\cite{Head2003,Wihelm2003,picu2011mechanics,Broedersz2014}. The energy of the network can be written as sum of bending and stretching energy, governed by the stretch rigidity $\mu$ and the bending rigidity $\kappa$. Networks with coordination number or connectivity $Z<2d$ exhibit a mechanical phase transition, which can be characterized by the network stiffness $K=\partial^2 E/\partial\gamma^2$, with $E$ being the network elastic energy at a given strain $\gamma$. For central force networks ($\kappa=0$), $K$ is discontinuous at a critical strain $\gamma=\gamma_c$, with $K=0$ for $\gamma<\gamma_c$ and $K\ge K_c>0$ for $\gamma\ge \gamma_c^+$~\cite{vermeulen2017geometry,merkel2019minimal,arzash2020finite,lerner2022scaling}, see Fig.~\ref{fig_1}(a). Above the critical point, the networks stiffen with $K-K_c\sim\mu \Delta \gamma^f$, where $\Delta \gamma=\gamma-\gamma_c$ is the relative strain with respect to $\gamma_c$.  For finite $\kappa$, the elasticity becomes continuous and $K\sim\kappa |\Delta \gamma|^{-\lambda}$ is observed for $\Delta \gamma<0$. The critical behavior can thus be described by two exponents $f$ and $\phi$, where $\phi=\lambda+f$ \cite{Sharma2016}.

\textit{Minimal model} \textendash
We introduce a minimal model in 2D that exhibits characteristic features of strain-controlled phase transition. Consider a chain of two connected segments, see Fig.~\ref{fig_1}(b).  Each segment is an elastic spring with stretch rigidity $\mu=1$ and rest length $\ell_0=1$. We deform the two ends of the chain in the horizontal direction $\hat{\bm e}_0$ with a reduced extensional strain $t=\epsilon-\epsilon_c$, which vanishes at the transition. By symmetry the middle node can move in the vertical direction, leading to an angle $\theta$ between each segment and $\hat{\bm e}_0$. In 2D there is a degeneracy in the sign of $\theta$. The Hamiltonian of the system is the total stretching energy, which is related to the extension of each segment, $\Delta \ell=(1+t)/\cos(\theta)-1$,
\begin{equation}
H_m=\Delta \ell^2\approx t^2 + t \theta^2+\theta^4/4\,.
\label{e1}
\end{equation} 
Here, we have kept the leading-order terms in $t$ and $\theta$. Equation (\ref{e1}) has a similar form as the Landau free energy~\cite{landau1936theory} with $t$ being a reduced-temperature-like parameter, and the minimum-energy solution is given by
\begin{equation}
\begin{aligned}
\theta=\left\{ 
\begin{aligned}
&\pm \sqrt{-2t}\qquad&(t<0)\\
&0  \qquad&(t\geq 0 )\\
\end{aligned}
\right.
\end{aligned} \, .
\label{e2}
\end{equation}
The system energy reads $E^{(0)}=t^2\Theta(t)$ with $\Theta$ being the Heaviside function, suggesting that the system has a critical point $t=0$: for $t<0$ the chain buckles, indicative of a floppy phase, while for $t>0$ the chain becomes straight and bears tension, corresponding to a rigid phase.  For simple shear, we expect $t\propto\Delta\gamma$ to lowest order. As discussed below, however, nonlinear corrections to this can become important for comparison with experiments. 

Next, we introduce the bending energy, which is modeled as harmonic energy that aligns segments to a particular angle $\theta_b$. The Hamiltonian then reads
\begin{equation}
\begin{aligned}
H_m&=t^2 +t \theta^2+\theta^4/4+\tilde\kappa(\theta-\theta_b)^2\\&\approx t^2 + t \theta^2+\theta^4/4-\kappa\theta\,,
\end{aligned}
\label{e3}
\end{equation} 
where we assume $\theta_b > 0$ without loss of generality and in view of the $Z_2$ symmetry. \SC{We assume $\kappa = 2\theta_b\tilde \kappa$ to be small.} In Eq.~(\ref{e3}) a constant term is neglected and only the leading term in $\theta$ is kept. An important assumption here is that $\theta$ in the relaxed state with respect to bending is in general different from $\theta$ in the critical state for $\kappa=0$. This $\kappa$ is similar to an external field in a Landau theory for ferromagnetism, but the alignment due to $\kappa$ is not global, in contrast with an external aligning field. 

Minimizing $H_m$ we get~\cite{Supp}
\begin{equation}
E^{(0)} = |t|^{2}{\cal E}^{(0)}_{\pm}(\kappa/|t|^{3/2})\,,
\label{e4}
\end{equation} 
which reflects the mean-field criticality of fiber networks. To demonstrate this, we note that $t$ is related to $\Delta\gamma$ by $t\propto\Delta\gamma+a(\Delta\gamma)^2$ to second order, where $a$ is of order unity and depends on the deformation mode~\footnote{Here the deformation mode of the network is not specified, and $\hat e_0$ is the direction of maximal extension.  For simple shear with small $\gamma_c$, $a=1/4$, see Ref.~\cite{Supp}.}. 
The differential modulus $K^{(0)}=\partial^2E^{(0)}/\partial\gamma^2$ is then
\begin{equation}
\begin{aligned}
K^{(0)} = |t|^{0}{\cal K}^{(0,0)}_{\pm}(\kappa/|t|^{3/2})+a |t|^{1}{\cal K}^{(0,1)}_{\pm}(\kappa/|t|^{3/2})\,,
\end{aligned}
\label{e5}
\end{equation} 
where the two scaling functions are found from the exact solution of the minimum energy state of Eq.~(\ref{e3}), see Fig.~\ref{fig_1} (c) and (d) \cite{Supp}. Equation (\ref{e5}) reproduces the mean-field criticality of fiber networks sketched in Fig.\ \ref{fig_1} (a): (i) In the rigid phase, $K^{(0)}-K_c\propto t^{f_0}$ for $\kappa=0$ with $f_0=1$ and $K_c=2$. (ii) In the floppy phase, $K^{(0)}$ vanishes for $\kappa=0$ and is given by $K^{(0)}\sim \kappa|t|^{-\lambda_0}$ for \SC{$\kappa\ll|t|^{\lambda_0}$}, where $\lambda_0=3/2$. These scaling exponents agree with previous mean-field derivations~\cite{Rens2018,merkel2019minimal,lerner2022scaling,chen2023effective}. Moreover, we find that the stiffness at the critical point is different for $\kappa=0$ and $\kappa\to0$, with a ratio $(\lim_{\kappa\to0}K)/K(\kappa=0)=1/3$. This difference and the scaling variable $\kappa/|t|^{3/2}$ are consistent with Ref.~\cite{lerner2022scaling}. 
Because the second term in Eq.~(\ref{e5}) contributes subdominantly to the elasticity, in what follows we assume $a=0$ and set $t=\Delta\gamma$ below.

\begin{figure}[b]
	\centering
	\includegraphics[scale=0.38]{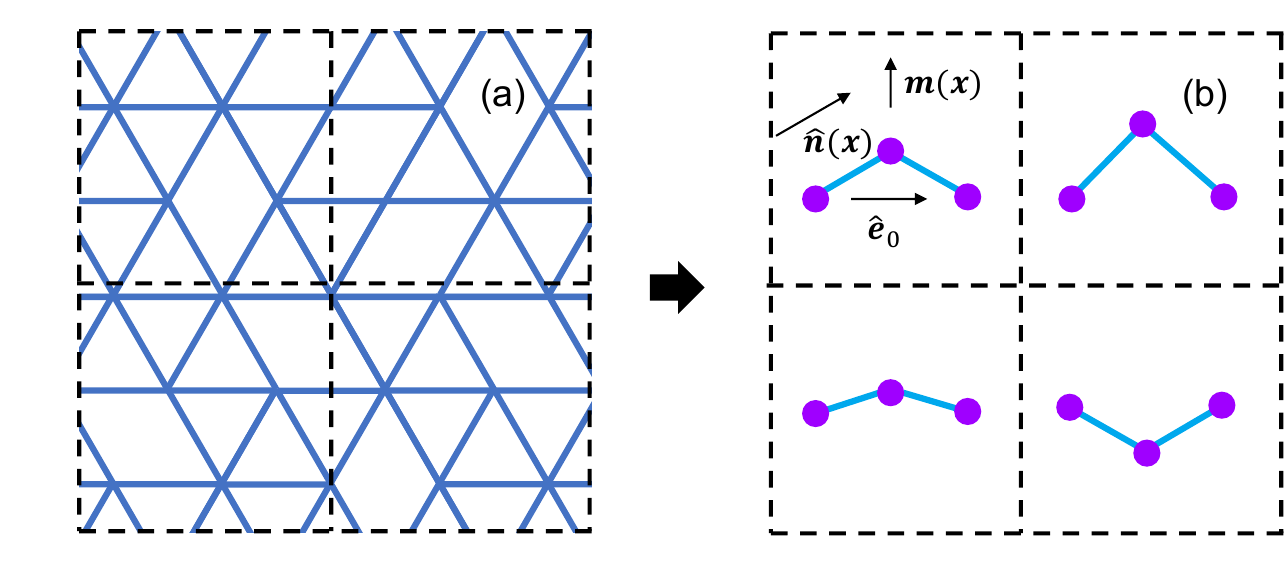}
	\caption{Schematic diagram of the coarse-graining process of the field theory. (a) A fiber network is divided into small blocks. (b) The network region in each block is approximated by a minimal model. 
	}
	\label{fig.rg}
\end{figure} 
\textit{Field theory} \textendash
We now extend the minimal model to a field theory.  Consider a fiber network in $d$ dimensions with system size $W$ and volume $V=W^d$, subject to an extensile strain  with relative strain $t$ in the direction ${\hat{\bm e}}_0$, representing an average direction of the developed force chain.  Here we do not specify the exact deformation mode of the network and neglect the deformation in the transverse direction(s) of $\hat{\bm e}_0$, see Ref.~\cite{Supp} for detailed discussion. We divide the network into small blocks and approximate each block with a minimal model, see Fig.~\ref{fig.rg}. The model parameter $\theta$ becomes a vector $\bm m = \hat{\bm n}-\hat{\bm e}_0$, with $\hat{\bm n}$ being the segment orientation of each minimal model.  

We start with the rigid phase, whose Hamiltonian reads $H=\int \diff \bm x \,h$ with
\begin{equation}
h=t^2 + t {\bm m}^2+{\bm m}^4/4
-\bm{\kappa}\cdot{\bm m}+A(\nabla {\bm m})^2+\bm c\cdot \bm m\,.
\label{e7}
\end{equation}
Here, $\bm m(\bm x)$ is a field that varies with the position $\bm x$ of the blocks.
We use a phenomenological interaction of the form \SC{$A(\nabla \bm m)^2=A(\nabla_im_j)^2$} that is lowest order in gradients and dominant at long wavelengths~\cite{deGennes1993,markovich2019shear}. This represents the elastic cost of the relative deformation between blocks. In Eq.\ \eqref{e7} we have also assumed that ${\bm m}$ is small near the critical point. Hence $\bm m = \hat{\bm n}-\hat{\bm e}_0$ is a $d-1$ dimensional transverse vector in the plane perpendicular to $\hat{\bm e}_0$, in which plane the effective bending rigidity ${\bm  \kappa}=\kappa \hat{\bm e}_1$ can also be assumed to lie, with $\hat{\bm e}_1$ being the direction of the bending force. $\bm c(\bm x)$ is a disordered field that is also transverse and emerges from the network quenched disorder, \SC{e.g., in the random crosslinking between fibers. $\bm c \cdot \bm m$ is the leading-order correction to the Hamiltonian due to such disorder, which is linear because disorder can locally break the rotational ($Z_2$ in 2D) symmetry of the minimal model.}  Importantly, both $A$ and $\bm c$ are functions of $\kappa$ and $t$, because they are phenomenological parameters that can change according to the network configuration. 

The minimum-energy state is found by minimizing $H$ with respect to ${\bm m}(\bm x)$. For an ordered network with translational symmetry (e.g., a perfect lattice), $\bm c = 0$. \SC{Because thermal fluctuations are not present in athermal fiber networks (unlike the Ising model), the disorder-free model} has a mean-field solution with an energy density $E/V=E^{(0)}$ ($V$ is the volume) and an elastic modulus $K/V = K^{(0)}$ as given above. Real networks, however, are disordered with nonzero $\bm c$, which leads to spatial fluctuations of $\bm m$ in the minimum-energy state. \SC{We postulate that the disorder modifies the energy to $E/V \simeq E^{(0)} + E^{(1)} $. Here $E^{(0)}$ assumes the same scaling form as Eq.~(\ref{e4}) (although the exact  scaling function may be different), and}
\begin{equation}
\begin{aligned}
	E^{(1)} = |t|^{2+f_1}{\cal E}^{(1)}_{\pm}(\kappa/|t|^{\phi_1}). 
\end{aligned}
\label{e13}
\end{equation} 
\SC{The exponent $2+f_1$ ensures $K-K_c\sim t^{f_1}$ in the rigid phase.} The complete elastic modulus reads $K/V\simeq K^{(0)}+K^{(1)}$, with 
\begin{equation}
\begin{aligned}
K^{(1)} = |t|^{f_1}{\cal K}^{(1)}_{\pm}(\kappa/|t|^{\phi_1}). 
\end{aligned}
\label{e14}
\end{equation} 
\SC{While Eq.~(\ref{e13}) is a scaling ansatz, the validity of this scaling form has been verified in prior simulations~\cite{Sharma2016,sharma2016strain,shivers2019scaling,arzash2020finite}. $E^{(1)}$ results from the combination of two terms: the spatial correlating term and the disorder term (last two terms of Eq.~(\ref{e7}), respectively). It vanishes in the absence of disorder ($\bm c \to0$) or when $A\to \infty$.}

The rigid phase is stretch-dominated, allowing us to take $\kappa=0$ for convenience. For small $t$ we expand  $A$ and $\bm c$ such that $A \simeq A_0$ and $\bm c(\bm x) \simeq \bm c_0(\bm x) +t\bm c_t(\bm x)$. The finite stretching elasticity in the rigid phase suggests that there is a non-zero interaction strength $A_0$. On the other hand, $\bm c_0$ must be zero to ensure a critical point at $t=0$, and the disordered field is characterized by $\bm c_t$ alone. For simplicity we take the disorder to be Gaussian and uncorrelated: $\langle  c_{t,\alpha}(\bm x) c_{t,\beta}(\bm x')\rangle=C_t^2\delta_{\alpha\beta}\delta(\bm x-\bm x')$. 

To obtain the scaling exponent $f_1$, we  expand Eq.~(\ref{e7}) to quadratic order in the fluctuations (Gaussian approximation)~\cite{kardar_2007}. We  extract the scaling exponent using standard methods~\cite{kardar_2007} and find $f_1 = d/2-1$ \cite{Supp}. For $d\geq 4$, we have $f_1\geq f_0$, such that $E^{(1)}$ cannot be distinguished from the mean-field behavior of $E^{(0)}$. Therefore, we predict the upper critical dimension $d_u=4$, see Ref.~\cite{Supp}. For $d=2$ and $d=3$, we expect a non-mean-field exponent $f=f_1$. While the exact value of this exponent is likely to differ from the Gaussian approximation due to higher order contributions from the fluctuations, it is important that $f_1<1$ in $d<4$, consistent with prior simulations~\cite{shivers2020nonlinear,arzash2020finite,arzash2021shear,arzash2022mechanics}. \SC{The full Hamiltonian including the fourth order term is numerically minimized in 2D, which gives $f_1\approx 0.43$~\cite{Supp}.} \SC{This is in reasonable agreement with previously reported exponents $0.34\sim0.55$ for bulk and uniaxial strain~\cite{shivers2020nonlinear,arzash2022mechanics}. Simulations find a
	dependence of the exponent on the mode of deformation and we do not
	expect our theory to accurately capture shear deformation since we only
	include deformation in a single direction $\hat{\bm e}_0$, see Ref.~\cite{Supp} for details.} In addition, a hyperscaling relation of $E^{(1)}$ yields a correlation length $\xi\sim |t|^{-\nu}$, where $\nu = (2+f_1)/d=1/d+1/2$~\cite{shivers2019scaling}. 

The floppy phase is more complicated. Because of the small or vanishing elasticity, the network can exhibit strong local strain fluctuations even in the absence of disorder. Therefore, we allow for the Hamiltonian density in Eq.\ \eqref{e7} to be modified by a fluctuating strain field $\tilde t(\bm x)$. 
To ensure a macroscopic relative strain $t$, we impose a set of constraints on $\tilde t$ for each $\cal L$, where $\cal L$ is any line that spans the network in the direction ${\hat{\bm e}}_0$:  $(1/W)\int_{\cal L}\diff s\,  \tilde t(\bm x(s))=t$, i.e., the average relative strain for each $\cal L$ is $t$. The network state is then found by minimization with respect to both $\tilde t$ and $\bm m$. 
We begin with central-force networks ($\kappa=0$), which has zero elasticity in the floppy phase. This suggests that no interaction exists between blocks in the floppy phase, leading to $A=0$.  In this case, the Hamiltonian has infinite number of degenerate ground states for $t<0$ if $\bm c=0$: As long as $2\tilde t (\bm x) +\bm m^2(\bm x)=0$ holds, the network has zero elastic energy. The deformation among these degenerate ground states corresponds to the floppy (zero) modes of real networks. Such a degeneracy vanishes for any non-zero $\bm c$, hence $\bm c$ must vanish for $\kappa=0$. 
This degeneracy also precludes any perturbation-based method in solving the minimum-energy state, and an analytical identification of the exponents $\lambda_1$ and $\phi_1$ is more challenging than in the rigid phase, see also Ref.~\cite{Supp}. For finite $\kappa$, an expansion in $\kappa$ leads to $A=\kappa A_1$ and $\bm c(\bm x)=\kappa\bm c_\kappa(\bm x)$, where $\bm c_\kappa$ is assumed to be Gaussian and uncorrelated with $\langle  c_{\kappa,\alpha}(\bm x) c_{\kappa,\beta}(\bm x')\rangle=C_\kappa^2\delta_{\alpha\beta}\delta(\bm x-\bm x')$. We numerically minimize the Hamiltonian in $d=2$ and extract a non-mean-field $\lambda_1 = 1.88$~\cite{Supp}. \SC{This is consistent with the values $1.85\sim 1.98$ obtained in previous simulations of 2D networks~\cite{arzash2020finite,shivers2020nonlinear,arzash2022mechanics}.} 

\begin{figure}[t]
	\centering
	\includegraphics[scale=0.42]{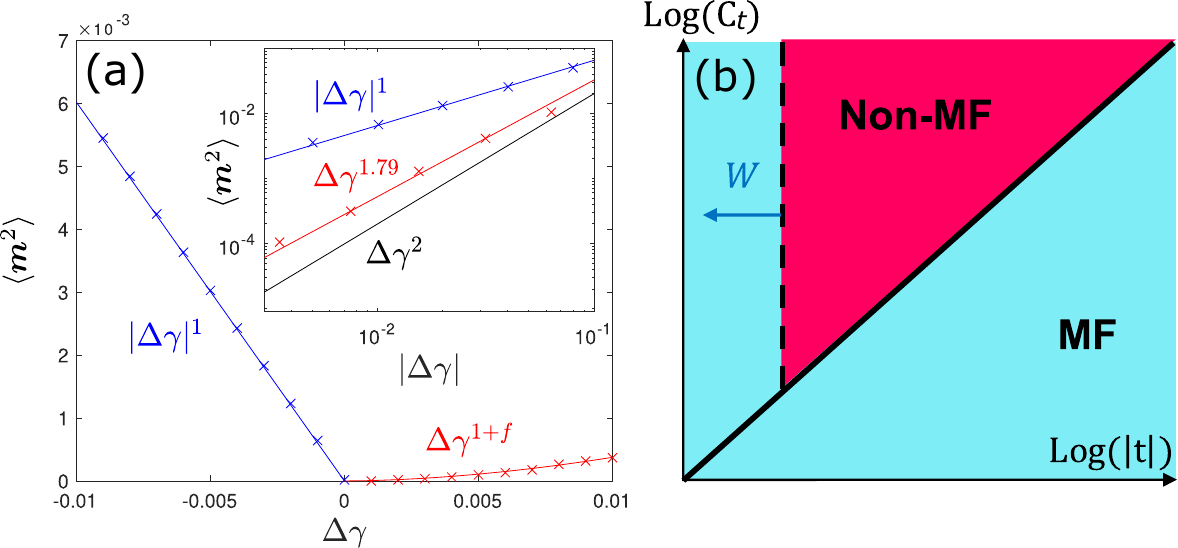}
	\caption{(a) Simulation results (crosses) of the  order parameter $\langle \bm m ^2\rangle$ for 2D central-force diluted triangular networks with connectivity $Z=3.3$ and system size $W=40$, subject to simple shear strain.   Theoretical fitting according to Eq.~(\ref{e18}) is shown in lines. The value of $f$ is obtained from Ref.~\cite{arzash2020finite}, $f=0.79$. Inset:  Scaling dependences of $\langle \bm m ^2\rangle$ for $\Delta \gamma<0$ (blue) and $\Delta \gamma>0$ (red). (b) Mean-field/non-mean-field diagram as function of the disorder $C_t$ and the absolute relative strain  $|t|$. The solid line indicates the Ginzburg criterion $|t|=t_G$ and the dashed line is the finite size criterion $|t|=t_W$, which decreases for increasing system size $W$, see Ref.~\cite{Supp} for further details. 
	}
	\label{fig.order}
\end{figure}

\textit{Order parameter} \textendash
An important quantity that has been missing in previous studies of fiber networks is the order parameter. Due to the local rotational symmetry of $\bm m$ ($Z_2$ symmetry in 2D) when $\kappa=0$, we do not expect a magnetization-like order parameter $\langle \bm m \rangle$. Instead, we consider $\langle \bm m ^2\rangle$ as the order parameter, which behaves as~\cite{Supp}
\begin{equation}
\begin{aligned}
\langle \bm m^2\rangle\sim\left\{ 
\begin{aligned}
&|t|^{1}\qquad&(t<0)\,\,\,\\
&t^{1+f}\qquad&(t>0)\,.\\
\end{aligned}
\right.
\end{aligned} 
\label{e18}
\end{equation}
Note that the scaling for $t<0$ can also be found for $\langle \theta^2\rangle$ of the minimal model.  For real networks, we construct an order parameter with a similar form, $\langle \bm m^2\rangle=\langle (\hat{\bm n}-{\hat{\bm n}}_c)^2\rangle_{\tau_c}$. Here, $\hat{\bm n}$ is the orientation of each real segment at a given strain. ${\hat{\bm n}}_c$ is the segment orientation at the critical strain. The average is taken with respect to all segments in a network with weight $\tau_c$. This weight is introduced because not all segments are involved in the phase transition: The segments that do not bear tension at the critical strain have no contribution to the elasticity. In Fig.~\ref{fig.order}(a) we show numerical results of $\langle \bm m^2\rangle$ for 2D diluted triangular lattices, which are in excellent agreement with Eq.~(\ref{e18}). While the nematic tensor $\uubar{\bm Q}=\langle {\hat{\bm n}}{\hat{\bm n}}-\uubar{\bm I}/d\rangle$ has been used to characterize nonlinear stiffening of fiber networks~\cite{feng2015alignment}, we find that $\uubar{\bm Q}$ is featureless at the transition for fiber networks, as shown in Fig. S3~\cite{Supp}. 

\textit{Discussion-}
Although the mathematical form of the model we present is reminiscent of a Ginzburg-Landau theory, a crucial distinction lies in the role of temperature: In the Ginzburg-Landau theory thermal fluctuations play a crucial role, while for athermal fiber networks such thermal fluctuations are absent. The non-mean-field behavior in our model arises from quenched network disorder. While we analyze the effects of small uncorrelated disorder, disorder in real networks can be more complicated. Specific correlations in the disorder can alter critical exponents, as is the case in disordered Ising models~\cite{kazmin2022critical}. Thus, it is not surprising that the exponents reported in previous numerical simulations depend on both network structure and deformation type, such that a universality class may not exist for fiber networks. 

Our work suggests that the upper critical dimension for fiber networks is $d_u=4$, in contrast with numerical evidence of $d_u=2$ for jamming, a superficially similar disordered rigidity transition. This difference may be due to differences in the nature of the disorder. Fiber networks have quenched permanent disorder that arises from the random cross-linking during network formation, while, the disorder in jamming is history-dependent and the local coordination is not fixed. It has been argued that the long wavelength disorder in jamming is suppressed close to the critical point, perhaps resulting in hyperuniformity~\cite{atkinson2016critical}. To explore the effects of hyperuniformity in our system, we have calculated the exponent $f$ in our model with hyperuniform disorder $\bm c(\bm x)$, and find that it indeed reduces the upper critical dimension~\cite{Supp}. This may help to explain the difference between fiber networks and jamming.

Our results show that the energy and resulting mechanical quantities cannot strictly be described by a single scaling function \cite{shivers2019scaling,lerner2022scaling}. However, the analysis above suggests that a recent scaling ansatz for the stiffness $K-K_c\Theta(t)\sim|t|^{f}{\cal K}_{\pm}(\kappa/|t|^{\phi})$ involving a single scaling function and constant $K_c$ in the rigid phase \cite{arzash2020finite} should be accurate for most of the simulations to date~\footnote{This is because simulations to date have been limited by finite-size
	effects for $|t|\lesssim 10^{-2}$ 
	and critical exponents are most accurately estimated for $\kappa\lesssim
	10^{-3}$. Thus, $\kappa/|t|^{3/2} \lesssim 1$ and $K^{(0)}$ can be approximated as a
	constant, with $K^{(1)}$ responsible for the stiffening, consistent with
	prior reports of non-mean-field behavior. Importantly, since the
	exponent $\phi$ has been reported to be larger than 2, it is still possible
	to observe the critical regime for $K^{(1)}$ with $K^{(0)}$ approximately constant.}. 
Precisely at, or very close to the critical point, the mean-field $E^{(0)}$ and $K^{(0)}$ should be observed due to finite size effects,  and these are consistent with recent scaling arguments for the critical point, as reported in Ref.\ \cite{lerner2022scaling}. 
Non-mean-field effects should only be observed within a certain range $t_W\lesssim|t|\lesssim t_G$, where $t_W\simeq W^{-1/\nu}$ represents the onset of finite-size effects where the correlation length becomes capped at the system size, see Fig.~\ref{fig.order}(b). 
The upper bound is derived using the Ginzburg criterion, $t_G\approx C_t^{4/(4-d)}L^{-2d/(4-d)}$, where $L$ is the average fiber length, see Ref.~\cite{Supp} for details.  

The model here provides a theoretical framework for critical phenomena in biopolymer networks. While we focus on static criticality in athermal networks, extensions to thermal~\cite{dennison2013fluctuation,mao2015mechanical,arzash2023mechanical}, dynamical~\cite{shivers2022nonaffinity} and possibly active~\cite{Chen2020,chen2022motor} networks should be addressable with standard methods~\cite{hohenberg1977,fruchart2021non}. Moreover, due to the similarity between the criticality of fiber networks and jamming, it would be interesting to explore whether a similar field theory can be constructed for jamming.

\begin{acknowledgements}
	\emph{Acknowledgments}:
	The authors acknowledge helpful discussions with M.E. Cates, M. Kardar, A. Liu and A. Sharma. The authors thank  the referees for insightful comments.
	This work was supported in part by the National
	Science Foundation Division of Materials Research
	(Grant No. DMR-2224030) and the National Science
	Foundation Center for Theoretical Biological Physics
	(Grant No. PHY-2019745). 
\end{acknowledgements}

\bibliography{citation}

\end{document}


\title{Supplementary Material \\ Field theory for mechanical criticality in disordered fiber networks}
\author{Sihan Chen$^{1,2,3,4}$, Tomer Markovich$^{2,5,6}$, Fred C. MacKintosh$^{1,2,7,8,9}$}
\affiliation{$^1$Department of Physics and Astronomy, Rice University, Houston, TX 77005, US\\
	$^2$Center for Theoretical Biological Physics, Rice University, Houston, TX 77005, US\\
	$^3$Kadanoff Center for Theoretical Physics, University of Chicago, Chicago, IL 60637, US\\
	$^4$The James Franck Institute, University of Chicago, Chicago, IL 60637, US\\
	$^5$School of Mechanical Engineering, Tel Aviv University, Tel Aviv 69978, Israel\\
	$^6$Center for Physics and Chemistry of Living Systems, Tel Aviv University, Tel Aviv 69978, Israel\\
	$^7$Department of Chemical and Biomolecular Engineering, Rice University, Houston, TX 77005\\
	$^8$Department of Chemistry, Rice University, Houston, TX 77005\\
	$^9$The Isaac Newton Institute for Mathematical Sciences, Cambridge University, Cambridge, UK}

\maketitle
	
\vspace{-1cm}

\section{Scaling functions of the minimal model}
	We derive here the scaling functions of the minimal model. We start with the Hamiltonian (Eq.~(3) of the main text)
	\begin{equation}
	\begin{aligned}
	H_m= t^2 + t \theta^2+\theta^4/4-\kappa\theta\,,
	\end{aligned}
	\label{S1.1}
	\end{equation} 
	Minimization of the Hamiltonian leads to an elastic energy
	\begin{equation}
	E^{(0)}(t,\kappa)=|t|^2{\cal E}^{(0)}_{\pm}(x_0),
	\end{equation}
	where $x_0=\kappa/|t|^{3/2}$, with two scaling functions ${\cal E}^{(0)}_{+}$ and ${\cal E}^{(0)}_{-}$ for positive and negative $t$, respectively:
	\begin{equation}
	\begin{aligned}
	{\cal E}^{(0)}_{+}(x)=&&
	\frac{1}{576} \Bigg[\sqrt[3]{2} \sqrt[6]{3}
	\sqrt[3]{\sqrt{81 x^2+96}+9 x} \Bigg(4 \sqrt{3}
	\left(\sqrt[3]{2} \sqrt[6]{3} \sqrt{27 x^2+32}+4
	\sqrt[3]{\sqrt{81 x^2+96}+9 x}\right)\\
	&&+9 x
	\left(\left(\sqrt{27 x^2+32}-3 \sqrt{3} x\right)
	\sqrt[3]{\sqrt{81 x^2+96}+9 x}-12 \sqrt[3]{2}
	\sqrt[6]{3}\right)\Bigg)+192\Bigg]\\
	\end{aligned}
	\end{equation}
	and
	\begin{equation}\begin{aligned}
	{\cal E}^{(0)}_{-}(x)=&&
	1+\frac{\left(\sqrt[3]{2} \left(\sqrt{81 x^2-96}+9
		x\right)^{2/3}+4 \sqrt[3]{3}\right)^4}{144\ 6^{2/3}
		\left(\sqrt{81 x^2-96}+9
		x\right)^{4/3}}
	-\frac{\left(\sqrt[3]{2}
		\left(\sqrt{81 x^2-96}+9 x\right)^{2/3}+4
		\sqrt[3]{3}\right)^2}{6 \sqrt[3]{6} \left(\sqrt{81
			x^2-96}+9 x\right)^{2/3}}\\
	&&
	-\frac{x \left(\sqrt[3]{2}
		\left(\sqrt{81 x^2-96}+9 x\right)^{2/3}+4
		\sqrt[3]{3}\right)}{6^{2/3} \sqrt[3]{\sqrt{81
				x^2-96}+9 x}}.
	\end{aligned}\end{equation}
	We assume that the reduced extensional strain $t=\epsilon-\epsilon_c$ is related to the shear strain $\Delta \gamma$ of a real network by $t\propto\Delta \gamma+a\Delta \gamma^2$. As an example we consider a simple shear deformation of the $x-y$ plane in the $x$ direction. Let there be a segment oriented along the extensional axis with end-to-end vector $\bm r_0 =(\sqrt 2/2) \bm x+ (\sqrt 2/2) \bm y$ before deformation. At the critical strain $\gamma_c$, its end-to-end vector becomes $\bm r_c =(\sqrt 2/2)(1+\gamma_c) \bm x+ (\sqrt 2/2) \bm y$. For strain $\gamma=\gamma_c+\Delta \gamma$, the end-to-end vector changes to $\bm r =(\sqrt 2/2)(1+\gamma) \bm x+ (\sqrt 2/2) \bm y$. The reduced extensional strain is defined as $t=(|\bm r|-|\bm r_c|)/|\bm r_c|$, which to leading orders in $\Delta \gamma$ reads $t=b(\Delta \gamma+a \Delta \gamma^2)$, with $a=(1-2\gamma_c)/4$ and $b=1/2$ for small $\gamma_c$. Note that here the extensional strain is evaluated with the reference state being the critical state. One may also choose the undeformed state as the reference state, and the resulting $b$ value would differ to first order in $\gamma_c$.  In the following we assume $b=1$ for simplicity such that $t\propto\Delta \gamma+a\Delta \gamma^2$. The stress is defined as $\sigma=\partial E/\partial \gamma$,
	\be
	\sigma
	=|t|\Sigma^{(0)}_\pm(x_0)\frac{\partial t}{\partial\gamma},
	\ee
	where 
	\be
	\Sigma^{(0)}_\pm(x)=\pm\left(2{\cal E}^{(0)}_{\pm}(x)-\frac{3}{2}x{\cal E'}^{(0)}_{\pm}(x)\right)
	\ee
	and
	\be
	\frac{\partial t}{\partial\gamma}=1+2a\Delta\gamma.
	\ee
	Note that $\Sigma^{(0)}_\pm(x_0)\ge0$.
	Also, note that for simple shear deformation $a<1/4$ for a segment oriented along the extensional axis. 
	The stiffness
	\be
	K^{(0)}=\frac{\partial\sigma}{\partial\gamma}=
	{\cal K}^{(0,0)}_\pm(x_0)
	\left(\frac{\partial t}{\partial\gamma}\right)^2
	+|t|\Sigma^{(0)}_\pm(x_0)\frac{\partial^2 t}{\partial\gamma^2}
	\simeq{\cal K}^{(0,0)}_\pm(x_0)+a|\Delta\gamma|{\cal K}^{(0,1)}_\pm(x_0),\label{Kmf}
	\ee
	where
	\be
	{\cal K}^{(0,0)}_\pm(x)=\pm\left({\Sigma}^{(0)}_{\pm}(x)-\frac{3}{2}x{\Sigma'}^{(0)}_{\pm}(x)\right)
	=2{\cal E}^{(0)}_{\pm}(x)-\frac{9}{4}x{{\cal E}'}^{(0)}_{\pm}(x)+\frac{9}{4}x^2{{\cal E}''}^{(0)}_{\pm}(x),
	\ee
	and
	\be
	{\cal K}^{(0,1)}_\pm(x)=6\left({\Sigma}^{(0)}_{\pm}(x)-x{\Sigma'}^{(0)}_{\pm}(x)\right)
	=\pm3\left(4{\cal E}^{(0)}_{\pm}(x)-4x{{\cal E}'}^{(0)}_{\pm}(x)+3x^2{{\cal E}''}^{(0)}_{\pm}(x)\right)
	\ee
	to first order in $\Delta\gamma$. Because $t\sim\Delta \gamma$ to first order in $\Delta \gamma$, Eq.~(\ref{Kmf}) reduces to Eq.~(5) of the main text.
	
\begin{figure}[t]
	\centering
	\includegraphics[scale=0.6]{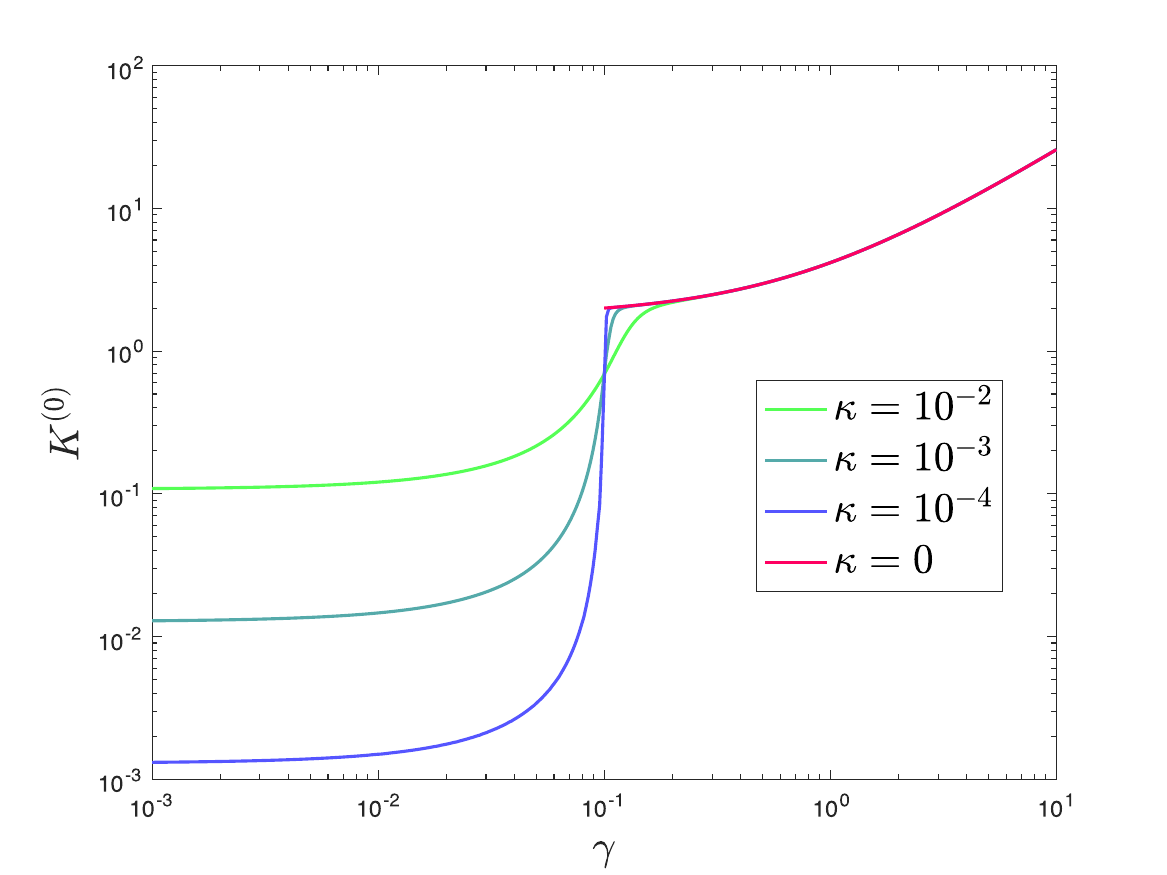}
	\caption{Example of the elasticity of the minimal model, calculated from Eq.~(\ref{Kmf}). Parameters used are: $a=0.2$ and $\gamma_c=0.1$. 
	}
	\label{fig.Kmf}
\end{figure}

	Note that Eq.\ \eqref{Kmf} means there is no single scaling function of the Widom form. This should only be expected approximately for $t\to0$. In Fig.~\ref{fig.Kmf} we plot the predicted $K^{(0)}$ as function of the strain $\gamma$, which reproduces the characteristic features the transition observed in prior experiments and simulations~\cite{Sharma2016}. 

\section{Scaling Exponents of the Field Theory}
\subsection{Rigid Phase}
We give here the detailed derivation of the scaling exponent $f_1$ in the rigid phase. Consider a network in a $d$-dimension space spanned by the base vectors $\{\hat{\bm e}_0$, $\hat{\bm e}_1$,...,$\hat{\bm e}_{d-1}\}$ and write the network Hamiltonian as $H=\int \diff^{d}\bm x\,h$, where $h$ is the Hamiltonian density (Eq.~(6) of the main text),
\begin{equation}
\begin{aligned}
h &= t^2+t {\bm m}^2+{\bm m}^4/4  -\bm\kappa\cdot{\bm m}+A(\nabla {\bm m})^2+\bm c\cdot \bm m\,.
\end{aligned}
\label{S00}
\end{equation}
As discussed in the main text, because the rigid phase is governed by the stretching energy, we may consider the simple case $\kappa=0$ only, in which $A\simeq A_0$ and $\bm c(\bm x) \simeq t \bm c_t(\bm x)$, leading to
\begin{equation}
\begin{aligned}
h &= t^2+t {\bm m}^2+{\bm m}^4/4  +A_0(\nabla {\bm m})^2+t\bm c_t\cdot \bm m\,.
\end{aligned}
\label{S000}
\end{equation}

We then perform a Fourier transform of Eq.~(\ref{S000}) with $\bm m(\bm x)=\sum_{j=1}^{d-1}\sum_{\bm q}m_{j,{\bm q}}\exp(i{\bm q}\cdot\bm x)\hat{\bm e}_j$ and $\bm c_t(\bm x)=\sum_{j=1}^{d-1}\sum_{\bm q}c_{t,j,{\bm q}}\exp(i{\bm q}\cdot\bm x)\hat{\bm e}_j$, while neglecting the $\bm m^4$ term as $t>0$. The Hamiltonian is rewritten as
\begin{equation}
\begin{aligned}
H/V &=t^2+\sum_{j=1}^{d-1}\sum_{\bm q}\left[(A_0q^2+t)|m_{j,{\bm q}}|^2+tc_{t,j,-{\bm q}}m_{j,{\bm q}}+t c_{t,j,{\bm q}}m_{j,-{\bm q}}\right]\,.
\end{aligned}
\label{S08}
\end{equation}
Minimizing $H$ with respect to $\bm m$ gives
\begin{equation}
\begin{aligned}
m_{j,{\bm q}}=\frac{-tc_{t,j,{\bm q}}}{A_0q^2+t}\,.
\end{aligned}
\label{S09}
\end{equation}

Substituting Eq.~(\ref{S09}) into Eq.~(\ref{S08}) gives the elastic energy density $E/V$. The average energy density is found by taking average with respect to the disorder $\bm c_t$ ($\langle c_{t,i,{\bm q}}c_{t,j,{\bm q'}}\rangle=(C_t^2/V)\delta_{i,j}\delta_{\bm q,-\bm q'}$):
\begin{equation}
\begin{aligned}
\langle E\rangle /V &=t^2-{(d-1)}\sum_{\bm q}\frac{\langle| c_{t,i,{\bm q}}|^2\rangle t^2}{A_0q^2+t}\\&=t^2-\frac{d-1}{V}\sum_{\bm q}\frac{C_t^2t^2}{A_0q^2+t}\\&=t^2-(d-1)\int  \frac{\diff^d{\bm q}}{(2\pi)^d}\,\frac{C_t^2t^2}{A_0q^2+t}\,.
\end{aligned}
\label{S15}
\end{equation}
\begin{table}[t]
	\begin{tabular}{|l|l|l|l|l|l|l|}
		\hline
		$d$ & \begin{tabular}[c]{@{}l@{}} Structure\end{tabular} & $Z$   & \begin{tabular}[c]{@{}l@{}}Deformation\end{tabular} & $f$    & $\lambda$    & Source  \\ \hline
		2         & TL                                                          & 3.3 & Shear                                                      & 0.79 & 1.85 & Ref.~\cite{arzash2020finite}  \\ \hline
		2         & PD                                                          & 3.3 & Shear                                                      & 0.85 & N/A  & Ref.~\cite{arzash2020finite} \\ \hline
		2         & PD                                                          & 3.2 & Uniaxial  Stress                                             & 0.55 & 1.95 & Ref.~\cite{shivers2020nonlinear} \\ \hline
		2         & TL                                                          & 3.3 & Bulk                                                       & 0.34 & 1.98 & Ref.~\cite{arzash2022mechanics}    \\ \hline
		3         & PD                                                          & 3.3 & Shear                                                      & 0.79 & 1.71 & Ref.~\cite{arzash2021shear}      \\ \hline
		3         & PD                                                          & 4.0 & Shear                                                      & 0.86 & 1.74 & Ref.~\cite{arzash2021shear}      \\ \hline
		3         & RGG                                                         & 3.3 & Shear                                                      & 0.92 & 1.88 & Ref.~\cite{arzash2021shear}     \\ \hline
	\end{tabular}
	\caption{Summary of non-mean-field exponents identified in previous numerical simulations of 2D and 3D networks. The network structures are: TL (triangular lattice), PD (jammed-packing-derived) and RGG (random geometric graph). }
	\label{exponents}
\end{table}
In the last equality we replace the sum with an integral in the large $V$ limit.  \SC{Notably, the integral in Eq.~(\ref{S15}) diverges for $d\geq 2$, which is dominated by the ultra-violet cutoff $q=1/\ell_0$, with $\ell_0=1$ being the length scale of fiber segment. For $d>2$, we rewrite Eq.~(\ref{S15}) with }
\SC{
\begin{equation}
\begin{aligned}
\langle E\rangle /V &=t^2-\Delta E + E^{(1)}\,,
\end{aligned}
\label{S22}
\end{equation}
where
\begin{equation}
\begin{aligned}
	\Delta E &= (d-1)\int_{q=0}^{1}  \frac{\diff^d{\bm q}}{(2\pi)^d}\,\frac{C_t^2t^2}{A_0q^2}\sim \frac{C_t^2}{A_0}t^2\,,\\
	E^{(1) }&= (d-1)\int  \frac{\diff^d{\bm q}}{(2\pi)^d}\,\frac{C_t^2t^3}{A_0q^2(q^2+t)}=(d-1)A_0^{-d/2}C_t^2t^{d/2+1}\int \frac{\diff^d \bm q' }{(2\pi)^d} \frac{1}{q'^2(q'^2+1)}\,.
\end{aligned}
\label{S23}
\end{equation}
In the last equality of Eq.~(\ref{S23}) we have used a variable substitution ${\bm q}' = \sqrt{A_0/t}{\bm q}$ to obtain the $t$-dependence. In Eq.~(\ref{S22}), $\Delta E\sim t^2$, which gives a correction to the mean-field part of the energy, $E^{(0)}$. The integral in $E^{(1)}$ converges for $d<4$, for which a non-mean-field exponent $f_1=d/2-1$ can be extracted. 
}
{The elastic modulus $K/V$ is calculated as  the second derivative of Eq.~(\ref{S15}) with respect to $t$, }
 \begin{equation}
 \begin{aligned}
K/V &=2-(d-1)\int  \frac{\diff^d{\bm q}}{(2\pi)^d}\,\frac{2C_t^2}{A_0q^2+t}\left[1-\frac{t}{A_0q^2+t}\right]^2\,.
 \end{aligned}
 \label{S19}
 \end{equation}
 \SC{We find that $K$ is strictly below the mean-field value $2$ for any non-zero disorder $C_t$. We then rewrite $K$ as
 \begin{equation}
 \begin{aligned}
 K/V&=2-\Delta K + K^{(1)}\,,
 \end{aligned}
 \label{S20}
 \end{equation}
where
\begin{equation}
\begin{aligned}
\Delta K &= 2(d-1)\int_{q=0}^{1}  \frac{\diff^d{\bm q}}{(2\pi)^d}\,\frac{C_t^2}{A_0q^2}\sim \frac{C_t^2}{A_0}\,,\\
K^{(1) }&= 2(d-1)A_0^{-d/2}C_t^2t^{d/2-1}\int \frac{\diff^d \bm q' }{(2\pi)^d} \left[\frac{1}{q'^2(q'^2+1)}+\frac{2}{(q'^2+1)^2}- \frac{1}{(q'^2+1)^3}\right]\,.
\end{aligned}
\label{S21}
\end{equation}
}
\SC{Here, $\Delta K$ gives  a correction to the mean-field elasticity $K^{(0)}$. It only modifies the magnitude of the discontinuity $K_c$ at $t=0$ and does not affect the exponent for $t>0$. The integral in $K^{(1)}$ converges for $d<4$, consistent with our expectation, $K^{(1)}\sim t^{f_1}$. }

\begin{figure}[t]
	\centering
	\includegraphics[scale=0.45]{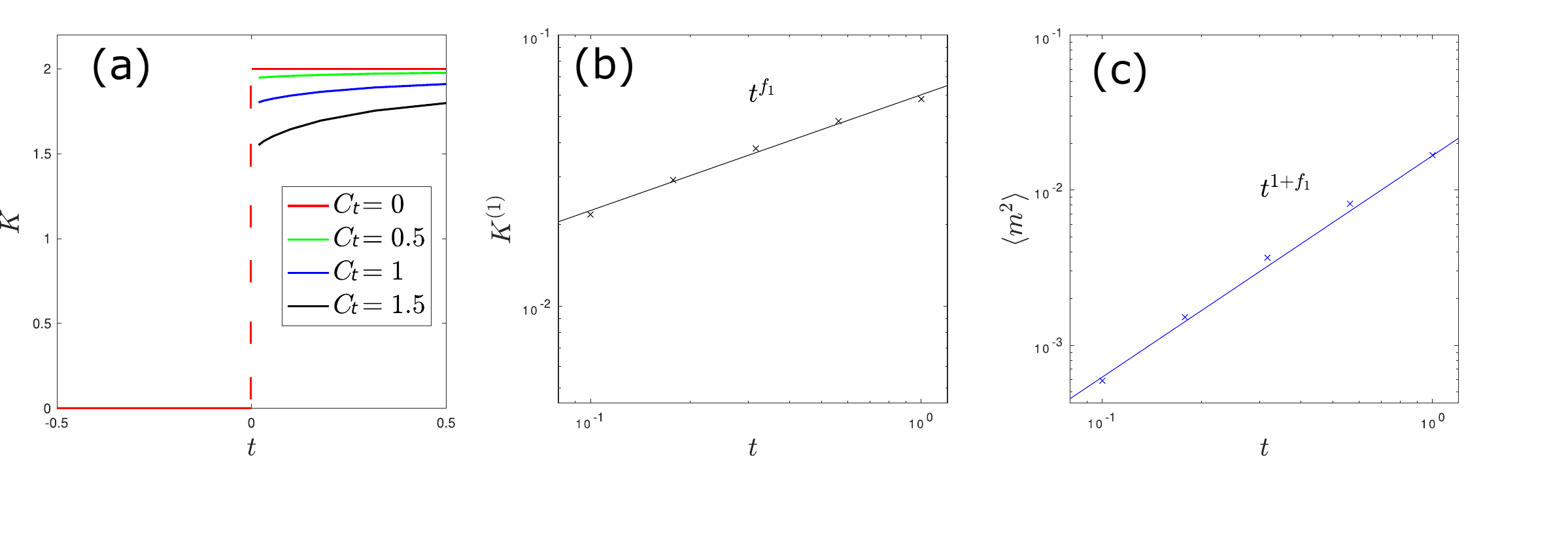}
	\caption{(a) Numerical solution of the elasticity in the rigid phase ($\kappa=0$), calculated from minimizing the Hamiltonian of Eq.~(\ref{S000}) with $A_0=1$ for various values of $C_t$. (b) Scaling of the numerical solution with $A_0=1$ and $C_t = 1$. An exponent $f_1\approx0.43$ is identified.  (c) Numerical results for $\langle {\bm m}^2\rangle$ in the rigid phase, fitted with an exponent $1+f_1$, where the value of $f_1$ is adopted from (b). 
	}
	\label{fig.numerical_rigid}
\end{figure}

\SC{The situation for $d=2$ is a bit more tricky because the dominant part of the integral in Eq.~(\ref{S15}) scales as $t^2 \log(t)$. This corresponds to a diverging term in the elasticity $\sim \log(t)$ for $t\to 0$.  However, in deriving Eq.~(\ref{S15}) the $\bm m^4$ term in Eq.~(\ref{S12}) is neglected. A numerical solution of the minimum energy state, which includes the $\bm m^4$ term, suggests that even in 2D, the qualitative features of $K$ are in alignment with our results for $d>2$: the discontinuity $K_c$ is reduced by the disorder, as shown in Fig.~\ref{fig.numerical_rigid} (a). The correction to $K_c$ increases for increasing $C_t$ as expected. A scaling dependence of $K^{(1)}=K-K_c$ is also identified, see Fig.~\ref{fig.numerical_rigid} (b).}

\SC{The scaling exponent $f_1$ governs the network stiffening in the rigid phase.} Because $f_1<f_0$ in $d<4$, we expect to observe non-mean-field exponents in dimensions $d=2$ and $d=3$. This agrees qualitatively with previous numerical results on 2D and 3D networks, see Table~\ref{exponents}. In the derivation above we have neglected the $\bm m^4$ term, whose presence will affect the value of the exponent. A numerical solution of the minimum-energy including the fourth order term in 2D gives $f_1\approx0.43$, see Fig.~\ref{fig.numerical_rigid} (b). Our numerical value of $f_1$ is close to the reported exponents of networks subject to uniaxial stress and bulk strain, but deviates from that of shear strain, see Table~\ref{exponents}. We speculate that the difference originates from the deformation in the transverse direction: In our calculation we neglect the deformation in the transverse direction, which best describes networks under uniaxial strain.  This may explain why the numerical value of $f_1$ is between that of bulk strain and uniaxial stress, since the bulk strain contains a transverse extension, and networks under uniaxial stress exhibit small transverse compression due to the Poissons' ratio~\cite{shivers2020nonlinear}. On the other hand, shear strain involves a large transverse compression due to volume preservation. In future work it would be interesting to explore the effects of  various deformation modes on the exponents. For this, we may need to modify the scalar $t$ to a second-rank tensor in order to describe an arbitrary deformation. 

\SC{The upper critical dimension $d_u=4$ can be found rigorously using the Ginzburg criterion (see Sec.~\ref{Ginzburg}) but it can also be found by asking when $f_1=1$, where we use the result of the Gaussian approximation, $f_1=d/2-1$.  For dimension $d>d_u$, because $f_1>1$, the disorder-contribution to $K-K_c$ is subdominant, and the mean-field contribution to $K-K_c$ dominates. However, non-mean-field behavior may still exist in other quantities for $d>d_u$. For example, the hyperscaling relation yields a correlation-length exponent $\nu=1/2+1/d$. Because fluctuations are absent in the mean-field limit (there are no thermal fluctuations), the correlation length is solely determined by the disorder in all dimensions. Hence, it is possible that a non-mean-field $\nu$ can be identified even for $d>4$ (the mean-field value of $\nu$ corresponds to its value for infinite $d$, i.e., $\nu_{\rm MF}=1/2$). }

In the rigid phase we set a uniform relative strain $t$ for all blocks of the network. Such uniformity is not a constraint imposed by the network structure, but is a consequence of the stress distribution in the rigid phase. In fact, as discussed in the main text, in the floppy phase there can be strong strain fluctuations. Below we detail the reason these fluctuations do not exist in the rigid phase. For this aim, we first relax the assumption of uniform relative strain, and introduce a relative strain field $\tilde t(\bm x)$ as we do for the floppy phase. In this case Eq.~(\ref{S000}) is modified to
\begin{equation}
\begin{aligned}
h &= \tilde t^2+\tilde t {\bm m}^2+{\bm m}^4/4  +A_0(\nabla {\bm m})^2+t\bm c_t\cdot \bm m\,.
\end{aligned}
\label{S01}
\end{equation}
A set of constraints $(1/W)\int_{\cal L}\diff s\,  \tilde t(\bm x(s))=t$ is imposed for any  line $\cal L$ that spans the network in the direction ${\hat{\bm e}}_0$, in order to ensure a macroscopic average strain $t$. 

Let us start with the simple case $\bm c_t = 0$, which has the mean-field solution
\begin{equation}
\begin{aligned}
\tilde t(\bm x) &= t\\
\bm m(\bm x) &= 0\,.\\
\end{aligned}
\label{S03}
\end{equation}
For small disorder $\bm c_t$, we let $\tilde t(\bm x) =t +u(\bm x)$ and expand Eq.~(\ref{S01}) to quadratic order of $u$, $\bm c$ and $\bm m$, which leads to
\begin{equation}
\begin{aligned}
h &=t^2+2tu + u^2+t\bm m^2+A_0(\nabla {\bm m})^2+t\bm c_t \cdot \bm m\,.
\end{aligned}
\label{S04}
\end{equation}
with a set of constraints $\int_{\cal L}\diff s\,  u(\bm x(s))=0$ for any line $\cal L$ that spans the network in the direction ${\hat{\bm e}}_0$. Applying Lagrange multiplier method, minimizing the Hamiltonian under this set of constraints is equivalent to minimizing 
\begin{equation}
\begin{aligned}
H^*=\int \diff{\bm x}\,[h(\bm x)+M(\bm y)u(\bm x)]\,,
\end{aligned}
\label{S05}
\end{equation}
where $\bm y$ is the projection of $\bm x$ in the subspace formed by $\{\hat{\bm e}_1$, $\hat{\bm e}_2$,...,$\hat{\bm e}_{d-1}\}$, i.e., $\bm y=\bm x-(\bm x\cdot\hat{\bm e}_0 )\hat{\bm e}_0$. $M$ is the Lagrange multiplier. Letting $\delta H^*/\delta u=0$, we have
\begin{equation}
\begin{aligned}
M(\bm y )&= -2t\\
u(\bm x) &= 0\,.
\end{aligned}
\label{S06}
\end{equation}
Equation (\ref{S06}) suggests that $\tilde t(\bm x)=t$, i.e., there are no spatial fluctuations of the relative strain. This is a result of the stress balance in the network: In the rigid phase there is nonzero stress, and the balance of stress between different network regions lead to a uniform relative strain field.

\subsection{Floppy Phase}
We now discuss the floppy phase, where the Hamiltonian density is 
\begin{equation}
h=\tilde t^2 + \tilde t {\bm m}^2+{\bm m}^4/4
-\bm{\kappa}\cdot{\bm m}+A(\nabla {\bm m})^2+\bm c\cdot \bm m\,,
\label{S16}
\end{equation}
with $A\simeq\kappa A_1$ and $\bm c(\bm x)\simeq \kappa \bm c_\kappa(\bm x)$. As discussed in the main text, for $\kappa=0$ the network ground states must satisfy $2{\tilde t}(\bm x)+{\bm m}^2(\bm x)=0$. According to Eq.~(\ref{S16}), central-force networks have zero energy in the floppy phase. Because the interaction strength is also zero, the network is free to deform within all ground states, which corresponds to zero modes or floppy modes. Such energy degeneracy suggests infinite linear response, which prevents any perturbation-based method. 

For finite $\kappa$ values, the network loses its ground-state energy degeneracy and has a unique ground state. \SC{However, for different disorder field $\bm c_\kappa$ there is different ground state. The variation between different ground states does not vanish for small $\kappa$ because both the disorder field $\bm c$ and the interaction strength $A$ are of same order of magnitude (both $\sim \kappa$). It is still possible to construct a Gaussian theory by neglecting high order terms. However, due to the large fluctuations of the ground state,  the Gaussian theory, or any perturbation-based method, is not reliable. This is because a perturbation method requires a reference ground state that the actual ground states converge to.} Therefore, a perturbation expansion of the Hamiltonian is not suitable in the floppy phase, and an analytical derivation of the exponent $\lambda_1$ is yet to be achieved. 

To explore the non-mean-field behavior in the floppy phase, we numerically calculate the elasticity of Eq.~(\ref{S16}) in $d=2$. We observe a non-mean-field exponent $\lambda_1 = 1.88$, see Fig.~\ref{fig.nematic} (a). This result is very close to the values identified in previous simulations of 2D networks, see Table.~\ref{exponents}.

\section{Order Parameter}
In this section we  derive the scaling behavior of the order parameter $\langle \bm m^2\rangle$ for central force networks, and discuss the use of other order parameters. In the rigid phase ($t>0$), we have $\langle \bm m^2\rangle=\sum_{j=1}^{d-1}\sum_{\bm q}|m_{j,{\bm q}}|^2$. Together with Eq.~(\ref{S09}), we get
\begin{equation}
\begin{aligned}
\langle \bm m^2\rangle &= (d-1)\int \frac{\diff^d{\bm q}}{(2\pi)^d}\,\frac{C_t^2t^4}{(A_0q^2+t)^2}
\\&=(d-1)A_0^{-d/2}C_t^2t^{d/2}\int \frac{\diff^d \bm q' }{(2\pi)^d} \frac{1}{(q'^2+1)^2}\\&\sim t^{f_1+1}\,.
\end{aligned}
\label{S11}
\end{equation}
\SC{Note that the integral in Eq.~(\ref{S11}) converges for $d<4$, hence, non-mean-field behavior can be found in the order parameter.} While this scaling relation is derived for the Gaussian model only, our numerical solution suggests that the scaling relation holds in the presence of the $\bm m^4$ term, see Fig.~\ref{fig.numerical_rigid} (c).

\begin{figure}[t]
	\centering
	\includegraphics[scale=0.45]{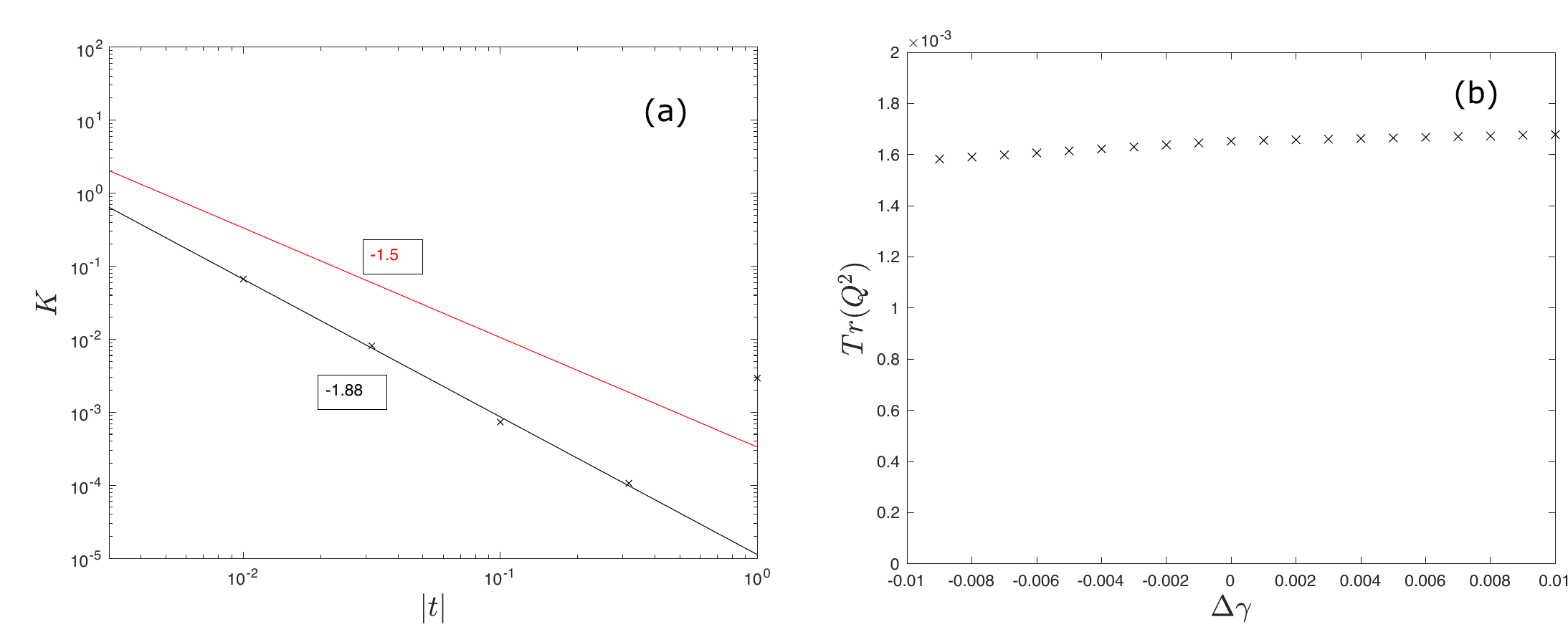}
	\caption{(a). Numerical solution of the elasticity in the floppy phase, calculated from minimizing the Hamiltonian of  Eq.~(\ref{S16}). Field theory parameters used in the calculation are: $\kappa=2\times10^{-4}$, $A_1 = 10$ and $B=1$. (b). Numerical results of ${\rm Tr}(\uubar{\bm Q})$, where $\uubar{\bm Q}$ is the nematic tensor, for 2D central-force diluted triangular networks with connectivity $Z=3.3$ and system size $W=40$. 
	}
	\label{fig.nematic}
\end{figure}
In the floppy phase, due to the degeneracy of the ground states for $\kappa=0$, $\bm m(\bm x) $ is not uniquely determined. However, as we discuss in the main text, all ground states must satisfy $2\tilde t(\bm x)+\bm m^2(\bm x)=0$. Because of the constraints $\int_{\cal L}\diff s\,  \tilde t(\bm x)=Wt$, we have $\langle \bm m^2\rangle=|t|/2$ for the field theory. 

For real networks, while the linear relation between $\langle \bm m^2\rangle$ and $|\Delta \gamma|$ is observed, we notice that the coefficient is slightly different from $1/2$.  This is possibly due to the fact that $t$ and $\Delta \gamma$ are not equivalent in real networks. For real networks, we expect $\tilde t=b\Delta\tilde \gamma$, where $\Delta\tilde \gamma $ is the local network strain. The coefficient $b$ may depend on the local network structure.

Because of the similarity between our theory and the Ising model, one may expect $\langle \bm m \rangle$ to be an order parameter. However, as we state in the main text, this is not an option because of the local rotational ($Z_2$ in 2D) symmetry of our model for $\kappa=0$, i.e., the energy is unchanged if $\bm m$ is rotated locally at any position in space.  Such local symmetry is absent in the Ising model because of the alignment interaction between adjacent spins, which corresponds to a non-zero value of the interaction strength $A$ in our model. Therefore, the Ising model only has a global rotational symmetry. Central force fiber networks ($\kappa=0$), on the other hand, has local rotational symmetry because $A=0$ in the floppy phase. For this reason, we choose $\langle \bm m^2\rangle$ as the order parameter, which remains unchanged under local rotation. 

Another orientation-related parameter is the nematic tensor $\uubar{\bm Q}=\langle {\hat{\bm n}}{\hat{\bm n}}-\uubar{\bm I}/d\rangle$, which is the order parameter for nematic liquid crystals~\cite{deGennes1993,chaikin1995}. Although the nematic tensor has been used to study the nonlinear stiffening of fiber networks~\cite{feng2015alignment}, it cannot serve as an order parameter for mechanical phase transition, because at the critical point the fiber segments still exhibit strong orientational variance. To prove this statement, we  calculated the nematic tensor  in simulations, and found that its value barely changes below and above the transition, see Fig.~\ref{fig.nematic} (b). Here the nematic tensor is defined as $\uubar{\bm Q}=\langle {\hat{\bm n}}{\hat{\bm n}}-\uubar{\bm I}/d\rangle_{\tau_c}$, where the average is weighted by the segment tension at the critical point. Because $\uubar{\bm Q}$ is a traceless tensor, we plot the trace of $\uubar{\bm Q}\cdot\uubar{\bm Q}$, which is the leading scalar quantity that can be constructed from $\uubar{\bm Q}$.

\section{Hyperuniformity}
We discuss here the effects of suppressed long-wavelength disorder. Specifically, we consider hyperuniform disorder that vanishes in the small $|\bm q|$ limit:
\begin{equation}
\begin{aligned}
\langle c_{t,i,{\bm q}}c_{t,j,{\bm q'}}\rangle=\delta_{i,j}\delta_{\bm q,-\bm q'}(C_t^2/V)|\bm q|^{\alpha}\,,
\end{aligned}
\label{S12}
\end{equation}
where $\alpha>0$. Substituting Eq.~(\ref{S12}) into Eqs.~(\ref{S15}, \ref{S20}) gives
\begin{equation}
\begin{aligned}
\langle E\rangle /V &=t^2-\Delta E + E^{(1)}
\end{aligned}
\label{S13}
\end{equation}
with
\begin{equation}
\begin{aligned}
\Delta E &= (d-1)\int_{q=0}^{1}  \frac{\diff^d{\bm q}}{(2\pi)^d}\,\frac{C_t^2t^2}{A_0q^{2-\alpha}}\,,\\
E^{(1) }&= (d-1)\int  \frac{\diff^d{\bm q}}{(2\pi)^d}\,\frac{C_t^2t^3}{A_0q^{2-\alpha}(q^2+t)}=(d-1)A_0^{-d/2}C_t^2t^{d/2+1}\int \frac{\diff^d \bm q' }{(2\pi)^d} \frac{1}{q'^{2-\alpha}(q'^2+1)}\,.
\end{aligned}
\label{S14}
\end{equation}
\SC{Here, the integral in $\Delta E$ diverges for $d> 2-\alpha$, contributing  a constant correction to the discontinuity in the elasticity. The integral in $E^{(1)}$ is finite for $d<4-\alpha$. 
The scaling exponent for this hyperuniform disorder is  $f_1=d/2-1+\alpha/2$, with an upper critical dimension $d_u=4-\alpha$.} This suggests that the upper critical dimension is strictly reduced by any hyperuniformity. In future work it would be interesting to computationally
test this prediction through hyperuniform dilution
of fiber networks.

\section{Ginzburg Criterion and Finite Size Effect}
\label{Ginzburg}
We derive here the 'Ginzburg criterion' of the theory~\cite{kardar_2007}, which indicates the range of $t$ for which non-mean-field behavior can be observed. For simplicity we consider $\kappa=0$, where the elasticity is zero in the floppy phase.  In the rigid phase, we have $K=K^{(0)}+K^{(1)}=K_c + 4a |t|+K^{(1)}$, where the prefactor $4$ is derived from the asymptotic limits of $K^{(0)}$. Therefore, $K^{(1)}\gg 4a|t|$ must be satisfied for the non-mean-field part of the elasticity to be observed. 

To estimate $K^{(1)}$, we write $E^{(1)}$ in terms of the correlation length $\xi$, $E^{(1)}\sim \xi^{-d}$. The correlation length diverges as $\xi\sim \xi_0|t|^{-(1/d+1/2)}$, where $\xi_0=C_t^{-2/d}A_0^{1/2}$.  $A_0^{1/2}$ is a characteristic lengthscale of the network. For randomly diluted networks, $A_0^{1/2}$ is approximated to be $L$, the average fiber length. With that, $K^{(1)}$ is found to be
\begin{equation}
\begin{aligned}
K^{(1)} &\sim \xi_0^{-d}|t|^{-1+d/2}\,.
\end{aligned}
\label{S17}
\end{equation}
The Ginzburg criterion of the field theory is then
\begin{equation}
\begin{aligned}
|t|\ll t_G\sim {\left(4\xi_0^da\right)^{-\frac{2}{4-d}}} \,.
\end{aligned}
\label{S18}
\end{equation}
Equation (\ref{S18}) is very similar to  the Ginzburg criterion of the Ginzburg Landau theory~\cite{kardar_2007}. Assuming the value of $a$ is of the order of unity,  the value of $\xi_0$ is the key parameter that affects the criterion. We estimate an upper bound of the relative strain, $t_G \approx C_t^{4/(4-d)}L^{-2d/(4-d)}$. This $L$-dependence may explain prior numerical results
showing the $f$-exponent approaches mean-field values for
increasing $L$, although this also coincided with the approach
to the isostatic point~\cite{sharma2016strain}. From Eq. (\ref{S18}) we further see that $d_u=4$ is the upper critical dimension, because Eq. (\ref{S18}) is never fulfilled for $d=4$. 

Another bound on $t$ comes from the finite size effect, i.e., in order to observe non-mean-field behavior, the correlation length $\xi$ needs to be smaller than the system size, leading to $|t|>t_W\sim W^{-1/\nu}$. Therefore, for any finite system size, mean-field behavior is expected for small enough $|t|$. Because of this finite size effect, prior simulations that found non-mean-field exponents have focused on $|t|\gtrsim 10^{-2}$. For these values of $|t|$, and for $\kappa\lesssim 10^{-3}$, one finds that $\kappa/|t|^{3/2}\lesssim 1$ and $K^{(0)}$ may be approximated as constant. On the other hand, because $\phi_1>3/2$, $\kappa/|t|^{\phi_1}$ can still be bigger than $1$, suggesting that $K^{(1)}$ dominates the stiffening. This explains prior results in which a single scaling function is sufficient to fit the simulation data~\cite{shivers2019scaling}.

\bibliography{citation}